# Design and Fabrication of Microfluidic Micro-Membranes by means of Two-Photon Polymerization

**Vladimir Osipov[a]\*, Alexander Dulebo[b] and David J. Webb[a]**

[a]Aston Institute of Photonic Technologies, Aston University, B4 7ET, Birmingham, UK
[b]Bruker Nano GmbH, Am Studio 2D, 12489 Berlin, Germany

**Abstract.** We show that with the use of a short wavelength (520nm) femtosecond laser source, Two-Photon Polymerization (2PP) fabrication of micro-membrane designs with channel sizes down to 1 µm is possible, using commercial photopolymers OrmoComp and FemtoBond. This approach, involving the flexible design and direct manufacturing of micro-filters and micro-membranes, can be applied for novel microfluidic devices.

\*Vladimir Osipov, v.osipov@aston.ac.uk

## 1 Introduction.

The modern field of microfluidics focuses on the precise control and manipulation of liquids in very small volumes. Microfluidic devices are especially important for lab-on-a-chip applications for biomedical, chemical or physical studies. To investigate phenomena occurring at a small scale, microfluidic channels are enhanced by the integration of even smaller elements, including active and passive components, such as filters, membranes, valves or mixers. However, conventional fabrication techniques for microfluidic chips have been largely limited to planar structures, holding back the exploitation of three-dimensional architectures, e.g. for multi-phase droplet separation or wet-phase fiber spinning Ref.1. Modern designs of microfluidic chips follow the "organ-on-chip" approach, which includes versatile 3D-positioning of microfluidic components, which are difficult to realize using conventional manufacturing techniques. Understanding the nature of the transport process and the precise evaluation of relevant parameters (especially size and porous 3D geometry) are of fundamental and practical interest Ref.2.

Polymeric membranes are widely used in various microfluidic chip designs, mainly due to their high performance. Thin membranes are particularly difficult to handle because the risk of tearing increases with the density and size of pores in the membrane. That is why many fabrication protocols involve molds fabricated of more solid material. However molds typically require pretreatment or the application of sacrificial layers to prevent membrane adhesion to the mold, which is usually hydrophilic Ref.3.

The main purpose of this work is to demonstrate the capability of the fs laser based 2PP technique for micro-membrane design and direct fabrication of channels with sizes down to 1 micron, enabling application in microfluidic systems for microparticle or microbiological object separation.

The 2PP laser fabrication technique, Ref.4, allows for direct fabrication of designed micro-scaffolds and porous-like micro-structures with micron or sub-micron scale precision: this is a radical and flexible concept that is not achievable by established additive manufacturing techniques and which leads to a range of novel device possibilities. For example, 2PP has been used to fabricate a set of compact optical components (lenses, waveguides, gratings etc) and can be used to create the micro-fluidic structure of a device incorporating precision mechanical components such as locks and shutters. It has previously been used for the fabrication of biomedical micro-optics for laser surgery Ref.5, three-focus intraocular lens design Ref.6, drug delivery Ref.7 and intra-vascular pressure sensing Ref.8.

Recently, 2PP lithography, was employed to print artificial channels of 4 μm in diameter Ref.9. The channel was printed into a traffic-cone-shaped structure including a needle shaped void for a traditional needle electrode to sit in. The needle was glued to the cone and then cast into an epoxy resin cube to form a needle-plane geometry with 2 mm insulation distance.

## 2 Methods

*2.1 Design*

We used COMSOL Multiphysics® 5.4 software to design our structures. COMSOL Multiphysics® is a simulation platform that provides fully coupled multi-physics and single-physics modelling capabilities. This platform includes all the steps in the modelling workflow — from defining geometries, material properties, and the physics that describe specific phenomena to performing computations and evaluating the results. Models deloped in COMSOL can be easily exported in stl-file format, that can be used for fabrication using a 3D printer or similar manufacturing device, such as our 2PP system.

*2.2. 2PP setup*

Aston Institute of Photonic Technologies has designed and built a custom 2PP system, which is based around a Chromacity Spark1040 laser (wavelength 520 nm, repetition rate 100 MHz, pulse length 120fs, average power 10 mW), LaserNanoFab galvano-scanner with controller and 3D nano-positioner with travel range 5mm, operated by M4D software. For the work described here, we used a GT Vision dry 60x objective with NA of 0.8. Scanning speed was 0.25 mm/s with slicing 500nm and hatching 250 nm. The ability to use 520 nm laser wavelength was chosen to enable a reduction in the sizes of fabricated microstructures in comparison with popular 780 nm laser radiation Ref.10. It has so far been tested on 3D micro-structure fabrication with SU-8, FemtoBond and Ormocomp photopolymers. The construction of the sample holder (Fig.1) allows 2PP fabrication processing of polymer films on both planar glass substrates or optical fibre-top end facet with precision about 200 nm.

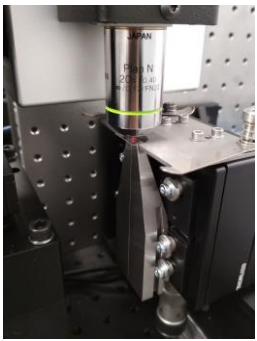

**Fig.1** Flexible 2PP fabrication facilities with possibility to process glass substrates or 125 µm diameter optical fiber end facets.

## 3 Results

The image of a proposed 3D design of micro-membrane with 95 x 95 x 5µm-size and with 10, 8, 6, 5, 4, 3, 2, 1 µm size square holes is shown in Fig.2a and an optical microscopy view of 2PP-fabricated FemtoBond polymer sample is shown in Fig.2b.

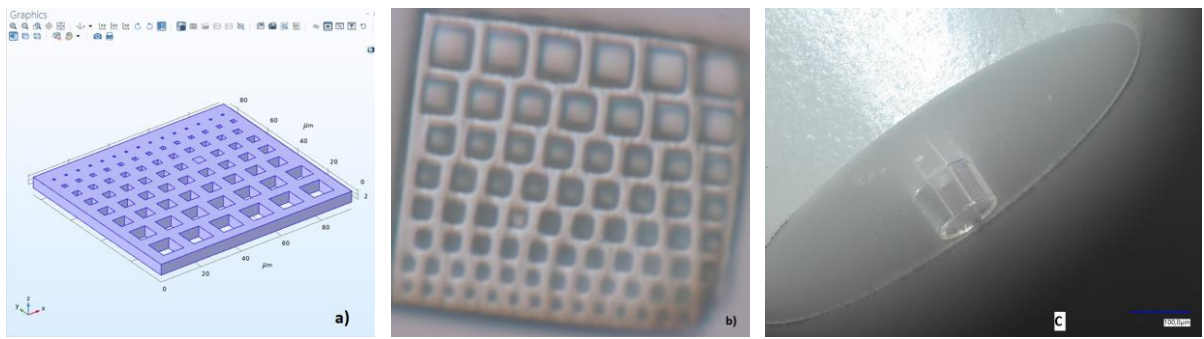

**Fig.2.** a) Proposed 3D design of micro-membrane; b) 95 x 95 x 5µm-size 2PP-fabricated FemtoBond sample with 10, 8, 6, 5, 4, 3, 2, 1µm size square holes on glass substrate, c) example of membrane-based sensor on optical fiber top in ferrule.

A possible realization of a micro-membrane based micro-sensor on an optical fiber endfacet, with the fiber embedded in a ceramic ferrule is demonstrated in Fig.2c. Here, the 2PP-fabricated microsensor diameter is 120 µm and the micro-membrane thickness is 3 µm (details will be published elsewhere).

## 4 AFM characterisation of the fabricated Micro-Membrane

The 2PP-fabricated sample of the micro-membrane shown in Fig.2b was studied using Bruker's Dimension Icon® Atomic Force Microscope (AFM) System.
Main parameters of the AFM System:
- Inspectable area 180x150x40 mm$^3$.
- Noise < 30 pm RMS
- 90x90x10 µm$^3$ scan size
- Imaging mode: PeakForce Tapping and Force Spectroscopy to measure stiffness
- Probes used: RTEPSA-300-125, RFESPA-75, SCANASYST-AIR-HPI –shown in Fig.3.

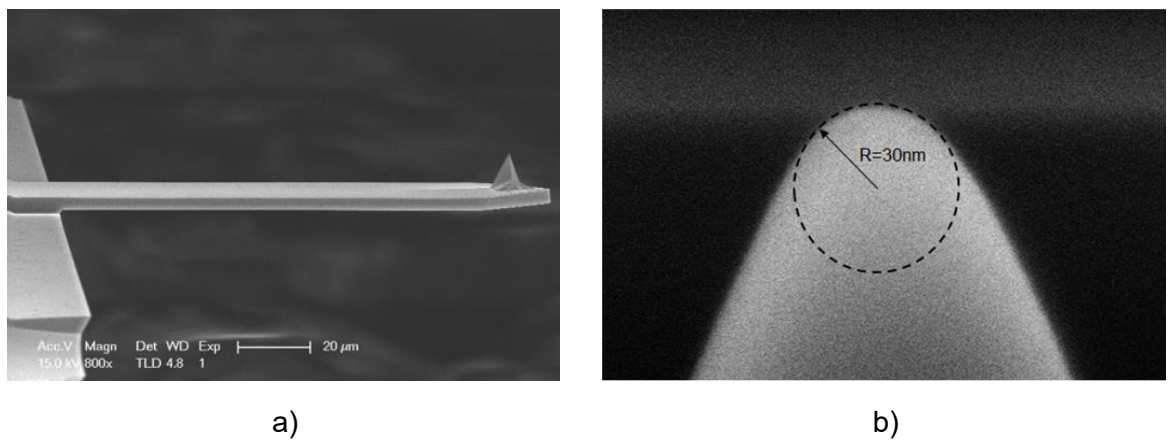

a)　　　　　　　　　　　　　　　　　　　　b)

**Fig.3.** The electron microscopy view of the probe (a) and its tip (b) for AFM testing of the fabricated sample.

The registered probe track position, cross section and 3D topography of the 2PP-fabricated micro-membrane are shown in Fig.4. It is clear that the realized membrane topography is in good correspondence to the designed one. The observed apparent reduction in hole size with depth can be explained by the conical form and geometrical size of the AFT-probe (Fig.3b).

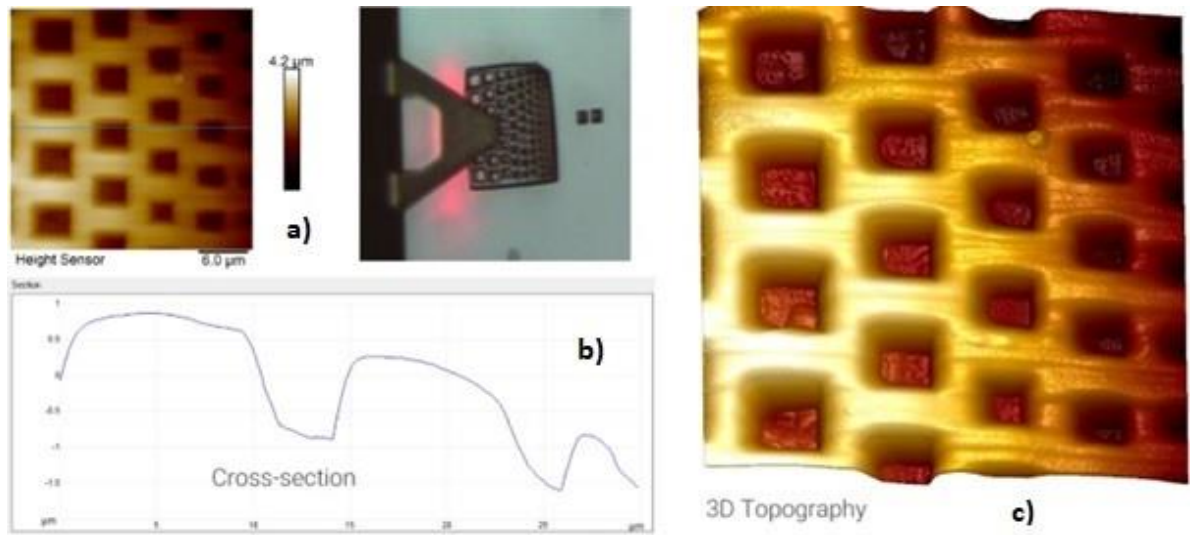

**Fig.4.** The registered probe track position (a), corresponding cross section (b) and 3D topography (c) of the 2PP-fabricated polymer micro-membrane. The blue line on the left photo (a) marks the probe track position on the sample, corresponding to the registered cross section on the lower graph.

## Conclusions

The proposed flexible design & fabrication approach is expected to be applied for prototyping of medical devices and sensors for biomedical engineering based on micro-filters and micro-membranes with sizes around 1 μm or less. The detailed characterization of the inner holes walls geometry needs specific methods like cylindric AFM probe application.

## Disclosures

The authors declare that there are no financial interests, commercial affiliations, or other potential conflicts of interest that could have influenced the objectivity of this research or the writing of this paper.

## Code and Data Availability

All datasets used in this work are publicly available through the corresponding references listed. In the meantime, specific inquiries can be directed to the corresponding author.

**Acknowledgements.**

We would like to gratefully acknowledge the LEVERHULME TRUST for support of this work, Research Project Grant RPG-2022-334.

**References.**

1. https://www.nanoscribe.com/en/news-insights/news/in-chip-printing-of-microfluidic-devices/

2. F. den Hoed, A. Ottomaniello, O. Tricinci, L. Ceseracciu, M. Carlotti, P. Raffa, and V. Mattoli, Facile Handling of 3D Two-Photon Polymerized Microstructures by Ultra-Conformable Freestanding Polymeric Membranes,  Adv. Funct. Mater. 2023, 33, 2214409.

3. K. Corral-Nájera, G. Chauhan, S. Serna-Saldívar, S. Martínez-Chapa and M. Aeinehvand, Polymeric and biological membranes for organ-ona-chip devices, Microsystems & Nanoengineering, 2023, 9, 107.

4. M. Farsari, B. Chichkov, Material processing: two-photon fabrication, Nature Photonics 3, 450–452 (2009).

5. V. Osipov, V. Pavelyev, D. Kachalov, A. Žukauskas, and B. Chichkov, Realization of binary radial diffractive optical elements by two-photon polymerization technique, Optics Express 2010, 18 (25), 25808-25814.

6. V. Osipov, L. L. Doskolovich, E. A. Bezus, W. Cheng, A. Gaidukeviciute, and B. Chichkov, Fabrication of three-focal diffractive lenses by two-photon polymerization technique, Appl Phys A, 2012, Issue 3, pp 525-529.

7. V. Pavelyev, V. Osipov, D. Kachalov, S. Khonina, W. Cheng, A. Gaidukeviciute, and B. Chichkov, Diffractive optical elements for the formation of "light bottle" intensity distributions, Applied Optics, Vol. 51, Issue 18, pp. 4215-4218 (2012).

8. R. K. Poduval, J. M. Coote, Ch. A. Mosse, M.C. Finlay, A. E. Desjardins, I. Papakonstantinou, Precision-Microfabricated Fiber-Optic Probe for Intravascular Pressure and Temperature Sensing, IEEE J. Selected Topics in Quant. Electronics, V.27, N. 4, 2021.

9. F. Liu, R. Saunders, S. Rowland, F. Aldawsari, Z. Luo, H. McDonald and S. Qi Li, Electrical Trees Grown from Micron-Scale Artificial Channels Fabricated by 2PP 3D Printing, 2023 IEEE Conference on Electrical Insulation and Dielectric Phenomena, DOI: 10.1109/CEIDP51414.2023.10410455.

10. https://support.nanoscribe.com/hc/en-gb/articles/360012401739-Technical-Parameters

**Vladimir Osipov** has been a researcher with Stepanov Institute of Physics (SIP), Belarus since 1976, Marie-Curie Fellow with the Laser Zentrum Hannover (LZH), Germany and Research Associate with Aston University, UK since 2017. He received his PhD in quantum electronics from SIP (1997). He has been awarded nine patents and has more than 75 technical publications. His interests include two-photon polymerization, micro-optics/micro-devices design and laser fabrication.

**Alexander Dulebo** is an application engineer with Bruker Nano, Inc., Germany since 2012. He received his PhD degree in biophysics from University of South Bohemia in České Budějovice, Czech Republic (2010). He has worked on various tasks, exploring life and materials at molecular, cellular and microscopic levels..

**David J. Webb** joined Aston University as Reader in Photonics in 2001, was promoted to Professor in 2012 and awarded a 50th Anniversary Chair in 2016. Prior to joining Aston he had spent 10 years as a Lecturer, then Senior Lecturer in the Physics Department at the University of Kent at Canterbury. He received BA in Physics (1982) at the University of Oxford and the PhD in Physics (1988) at the University of Kent. His professional work has primarily focused on solving complex object detection and tracking problems across a variety of domains, including overhead imagery, robotics, and real-time video analysis systems.